\documentclass[%
 reprint,
 amsmath,amssymb,
 aps,
pra,
 floats,
]{revtex4-1}

\usepackage{times}
\usepackage{amsmath}
\usepackage{graphicx}
\usepackage{subcaption}
\usepackage[T1]{fontenc}
\usepackage{amsfonts}
\usepackage{amssymb}
\usepackage{anyfontsize}
\usepackage{csquotes}
\usepackage{titlesec}
\usepackage{subcaption}

\begin{document}

\preprint{APS/123-QED}

\title{Tunable low energy $P_{s}$ beam for the anti-hydrogen free fall and for testing gravity with a Mach-Zehnder interferometer}

\author{Andre Rosowsky}
\affiliation{IRFU,CEA, Universit\'{e} Paris-Saclay}
\altaffiliation[Address: ]{CEN Saclay, F-91191 GIF-sur-YVETTES Cedex, FRANCE}
\email{andre.rosowsky@cern.ch}

\date{\today}

\begin{abstract}
The test of gravitational force on antimatter in the field of the matter gravitational field, produced by earth, can be done by a free fall experiment which involves only General Relativity, and with a Mach-Zehnder interferometer which involves Quantum Mechanics. This article presents a new method to produce a tunable low energy $(P_{s})$ beam suitable for trapping the $(\bar{H}^{+})$ ion in a free fall experiment, and suitable for a gravity Mach-Zehnder interferometer with $(P_{s})$. The low energy $(P_{s})$ beam is tunable in the [10 eV, 100 eV] range.

\begin{description}
\item[PACS numbers]$36.10.$Gv,$34.35.+$a,$ 68.49.$Bc,$41.75.$Fr,$ 41.75.$Ht,$73.20.$Mf,$34.80.$Pa,$03.75.$Hh,$ 03.75.$Kk
\item[keywords]Antimatter gravity,Antihydrogen,Positronium,Positron
\end{description}
\end{abstract}

\maketitle

\section{\label{sec:levelA1}Introduction}
When the Standard Model of particles interactions and General Relativity are involved in the same phenomena a number of questions are unanswered. Assuming the CPT invariance\cite{CPT-theorem}\cite{CPT-status}\cite{CPT-tests}, the theory states that anti-matter attracts anti-matter in the same way as matter attracts matter, but it tells nothing about the attraction, or the repulsion, of an anti-matter atom in the field of matter mass like the one generated by the earth. The Standard Model of particles interactions is mute on the subject since it does not include gravitation. Anti-matter is a quantum field theory concept which is not required by General Relativity. In fact a consistent Quantum Field Theory capable of merging General Relativity with the Standard Model has still to be found.

\vspace{2mm}
\par

When Quantum Field Theory and General Relativity are both considered to interpret an experiment, they lead to diverging points of views and creative controversies that will be presented in the frame of positronium gravitational interferometry.

\vspace{2mm}
\par
Even in the frame of pure General Relativity, the galaxy rotation curve cannot be explained, which lead to introduce dark matter and dark energy, that is new quantum fields. These dark quantities are there to compensate for the inconsistency of the curve with General Relativity, but are not required by the Standard Model which is experimentally confirmed at unprecendent precision.

\vspace{2mm}
\par
General relativity considers only one type of mass, and does not make any distinction for the parameter called mass when it is used for particles or for anti-particles. Mass in the frame of General Relativity is both the dynamical variable related to inertia via the energy-momentum, and to gravitation via the free fall along a geodesic. In Quantum field theory mass is only related to inertia via the energy-momentum, but not to gravitation.

\vspace{2mm}
\par

These situations have triggered a lot of efforts to perform a \textit{free fall} experiment with an atom of anti-hydrogen. Since there is no agreed prediction by theory, here \textit{free fall} of an atom of anti-hydrogen released at rest, means that it can go up or down. The main difficulty to perform the experiment is to produce an atom of anti-hydrogen with an horizontal speed of the order of $60 \; m.s^{-1}$ or less, and with a spread of vertical speed corresponding to a temperature below $10 \; \mu$eV. In a trap which contains a cloud of particles, the temperature refers to the kinetic energy spread in the particles kinetic energy distribution. But in the free-fall experiment there is only 1 particle involved, not a cloud. The temperature here, refers to the cooling cloud and hence to the atom residual uncontroled kinetic energy after cooling, that shall be below $10 \; \mu$eV. 

\vspace{2mm}
\par

The test of anti-hydrogen free fall is a General relativity experiment which does not involves Quantum Mechanics. Conversely, the Colella Overhauser Werner experiment (COW)\cite{cow1}\cite{cow2} performed in 1959, was a test of gravitational effect on the de Broglie wave length of neutrons : it involved both General Relativity, via gravitation, and Quantum Mechanics. The same test can be done with an atom made in equal parts of matter and anti-matter : positronium $(P_{s})$. If the free fall experiment finds that the matter anti-matter interaction is repulsive, then the interferometer experiment will measure this difference provided that General relativity and Quantum Mechanics are both valid. Furthermore, if the repulsive force intensity is equal to the attractive force between matter particles, then the interference oscillation pattern of the COW experiment shall vanish in the case of the $(P_{s})$ atom. Hence both experiments are tightly related. 

\vspace{2mm}
\par

In 1995, at CERN, the PS210 experiment produced 9 atoms of antihydrogen at the Low Energy Antiproton Ring (LEAR) with an antiproton $(p^{-})$ beam crossing a Xenon jet target. So the first experiment to produce $(\bar{H})$ atoms was a \textit{beam-on-target experiment}. The following experiments performed at the Antiproton Decelerator (AD) were all based on electromagnetic traps, hence \textit{bottle type experiment} : ATRAP\cite{ATRAP2}\cite{ATRAP1}, ATHENA\cite{ATHENA1}\cite{ATHENA2}, ASACUSA\cite{ASACUSA1}. The success of these experiments in producing $(\bar{H})$ in large quantities, or in the case of ASACUSA storing large amount of  $(p^{-})$ in a trap, opened the door to the first measurements on $(\bar{H})$. Since the trapped particles carry little kinetic energy, the key parameter in these experiments is the particle cloud temperature : a bottle experiment scientific potential is as great as the cloud temperature is low. In the first experiments, the temperature achieved was not low enough to allow for a free fall experiment. Also at the time, it was not commonly realized that neither the Standard Model, nor the CPT invariance, would make any prediction on the attractive or repulsive feature of the gravitational force between matter and anti-matter, or conversely would trigger controversies. The next generation of bottle type experiments, ALPHA\cite{ALPHA_1}\cite{ALPHA_G1}, BASE\cite{BASE1}, CUSP\cite{CUSP1}(CUSP is the follow-up of ASACUSA) aim at lower temperatures, which, in principle, will allow the ALPHA experiment to detect the attractive or repulsive feature of matter-antimatter gravitation, will allow CUSP to measure with accuracy the hyperfine transition in the ground state n=1, and will allow BASE to measure with accuracy the proton-antiproton charge to mass ratio and magnetic moment. None of these bottle type experiment is designed to directly observe and measure the free-fall of an atom of antihydrogen in the matter field of gravitation of earth.

\vspace{2mm}
\par

Conversely, two beam-on-target experiments, AEGIS\cite{AEGIS1} \cite{AEGIS2} and GBAR\cite{GBAR1}\cite{GBAR2} are designed specifically with the goal to observe and measure the free-fall, that can go up or down, of an antihydrogen atom.

\vspace{2mm}
\par

GBAR relies on the production of the ion of anti-matter $(\bar{H}^{+})$ in two steps.

\begin{equation} \label{eq:1}
\begin{aligned}
& p^{-}  +  P_s \rightarrow \bar{H}+ e^{-}
\end{aligned}
\end{equation} 

\begin{equation} \label{eq:2}
\begin{aligned}
&\bar{H}  +  P_s \rightarrow \bar{H} ^{+} + e^{-}
\end{aligned}
\end{equation}

So the first goal of this experiment is not to produce the atom $(\bar{H})$, but the ion $(\bar{H}^{+})$. Unlike the atom, the ion can be cooled at the temperature of $10 \; \mu$eV \cite{Waltz}. Both experiments have choosen a positronium target, that is a cloud of $(P_{s})$ atom at rest, with kinetic energy of $\sim$1 eV. The beam is made of a precooled cloud of antiprotons $(p^{-})$ stored in trap before being accelerated and packaged into a beam of few keV.

\vspace{2mm}
\par

The opposite configuration, that is a beam of $(P_{s})$ atoms on a $(p^{-})$ target at rest, requires a low energy beam of $(P_{s})$ atoms.

\vspace{2mm}
\par

Using a low energy $(P_{s})$ beam can be an alternative approach to the production scheme designed for GBAR\cite{rosowskyperez}. If implemented by GBAR, this method requires some modifications in its steps and layout. The beam of $(P_{s})$ atoms on a $(p^{-})$ target at rest could also provide a timing window to the CUSP hyperfine measurement. To compute the minimum number of  $(p^{-})$  and  $(e^{+})$ needed to perform an antihydrogen free fall experiment, an hypothetical conversion of CUSP will be discussed.  

\vspace{2mm}
\par

Once a $(P_{s})$ beam is available, the complementary gravity experiment, a gravity quantum mechanical experiment, becomes feasible. Such an experiment will be described: it is a positronium version of the Colella Overhauser Werner experiment (COW) that was performed with neutrons.

\vspace{2mm}
\par

The first section will present the production of the $(P_{s})$ beam. The second section will discuss the kinematic constraints of the beam on target reactions (\ref{eq:1}) and (\ref{eq:2}) seen from two reference frames: $(p^{-})$ beam on a $(P_{s})$ target, and $(P_{s})$ beam on a $(p^{-})$ target. An estimation of the number of ions $(\bar{H}^{+})$ that could be produced by CUSP and GBAR with this layout will be computed. The third section will present and discuss the gravity interferometry with a $(P_{s})$ beam.

\section{The $(P_{s})$ beam.}

The physics of $(P_{s})$ formation when a positron beam hits a metal target has been subject of several experiments and theoretical studies between 1970 and 1990. The main findings are :
\vspace{2mm}
\par

\begin{itemize}
\item $(P_{s})$ is not produced in the bulk but near the surface of the metal,
\item positrons with kinetic energy of few keV thermalise while migrating from the bulk to the surface where they are either trapped, or form $(P_{s})$ which is itself trapped, by vacancies or oxygen contamination,
\item metals which have a negative work function eject the $(P_{s})$ atoms and the positrons at very low kinetic energy corresponding to the negative work function value,
\item when the temperature of the metal target is raised, the $(P_{s})$ atoms trapped on the surface can free themselves more easily and the fraction of $(P_{s})$ ejected from the metal increases.  
\end{itemize} 
\vspace{2mm}
\par

Most of the theoretical and experimental effort was focused on thermalized positrons from an incident beam with kinetic energy of $\sim$1 keV or more. The role assigned to $(P_{s})$ atoms was to measure the void cavities under the skin of the material. We will concentrate here on a different channel : the scattering of positrons at low energy, between 10 eV and 100 eV, but above the surface trapping of $(e^{+} , P_{s})$ at lower positron incident energy.
\vspace{2mm}
\par

Below 1 keV, two channels are no more neglectable and compete with the thermal desorption of $(P_{s})$ created from thermalized positrons:

\begin{itemize}
\item like electrons, at about 1 keV incident kinetic energy, the number of positrons backscattered after few interactions becomes large,
\item below 100 eV the tail of the backscattered positrons overlaps with the $(e^{+} + e^{-} \rightarrow P_{s})$ cross section, producing backscattered $(P_{s})$ atoms.
\end{itemize}
\vspace{2mm}
\par

Since for many years the goal was to produce very low energy positrons and $(P_{s})$ sources, the effort was focused on thermalized positrons, that is positrons implemented in the metal skin with kinetic energies of 1 keV or more. In that perspective, backscattered $(P_{s})$ was a source of loss and noise, and was often disregarded. Conversely $(e^{+})$ diffraction and reflection were studied, and provide experimental data on $(P_{s})$ as a side effect. The article \cite{backscattered2} which presents the results of experimental Bragg reflection of $(e^{+})$ on Al and Cu notes that the reflection of $(P_{s})$ is constant for $(e^{+})$ incident energy between 0 eV and 40 eV. But there were very few measurements of backscattered $(P_{s})$ from a metal, and the one focused on this topic, by R. H. Howell , I. J. Rosenberg , M. J. Fluss, was at energies at or above 50 eV \cite{backscattered}. Nevertheless this precious data will be used to estimate the the backscattered fraction of $(P_{s})$ suitables to produce a low diverging beam. 

\vspace{2mm}
\par

Unlike thermalized positrons which produce $(P_{s})$ at low energy and in a wide angular cone around the metal normal direction \cite{angulardistribution}, the backscattered $(P_{s})$ are emitted within a small angle, producing a $(P_{s})$ beam. 

\vspace{2mm}
\par

Few alternative paths to produce a $(P_{s})$ beam  were proposed. Photodetachement of the positronium negative ion was used to produce a tunable beam with energy in the [300 eV, 1.9 keV] window \cite{backscattered5a}. The beam is to high in energy and not coherent, hence not suitable for interferometry. Charge exchange reaction of energetic positrons with gas molecules is an other approach \cite{backscattered5b}\cite{backscattered5c}, but as for the photodetachement, the beam energy is to hight, the beam is not coherent and the efficiency is low.
\vspace{2mm}
\par

Conversely, scattering of positrons at glancing angle\cite{backscattered6} can be used to produce a coherent $(e^{+})$ beam. This approach with higher energies, $\sim20$ keV, is used to study crystal surface\cite{backscattered6a} : the reflection high-energy positron diffraction (RHEPD). But the high kinetic energy prevents the conversion of the $(e^{+})$ beam into a $(P_{s})$ beam and leads to low efficiency in converting $(e^{+})$ into $(P_{s})$ beam  ($\sim5\%$). 
\vspace{2mm}
\par

Compared to the above methods, the backscattered diffracted $(e^{+})$ approach proposed here, leads to higher conversion efficiency as osberved by R. H. Howell , I. J. Rosenberg , M. J. Fluss, and with the usefull feature of being coherent, and hence suitable for interferometry.

\subsection{Low energy positron diffraction from a metal crystal}

The measurements of the fraction of backscattered $(P_{s})$ from an incident $(e^{+})$ beam performed by R. H. Howell , I. J. Rosenberg, M. J. Fluss, are synthetized in table (\ref{table_1}). These values are visually extracted from the article, hence their precision is $\sim 1\%$. The graph build with these values corresponds to the figure 6 of their article. This experiment has a mask which limits the acceptance for the backscattered $(P_{s})$ within 30 degrees with respect to the normal incident beam.

\begin{table}[h]
\centering
\caption{Fraction of backscattered Ps within a cone of 1/2 opening angle 30 degrees}
\label{table_1}
\begin{tabular}{|l|l|l|l|l|}
\hline
 $K(e^{+}) \quad eV$ & Al & Cu & Au & Ni\\
\hline
50     & 0.42  & 0.26 & 0.20 & 0.22 \\
100   & 0.26  & 0.21 & 0.20 & 0.17 \\
200   & 0.15  & 0.16 & 0.17 & 0.13 \\
500   & 0.09  & 0.12 & 0.16 &    -   \\
1000 & 0.04  &    -   & 0.10 & 0.09 \\
2000 &   -     & 0.04 & 0.08 &     -  \\
3000 &   -     &   -    & 0.05 & 0.03 \\
\hline
\end{tabular}
\end{table}

\begin{figure}[htb]
  \centering
  \includegraphics[width=0.5\textwidth]{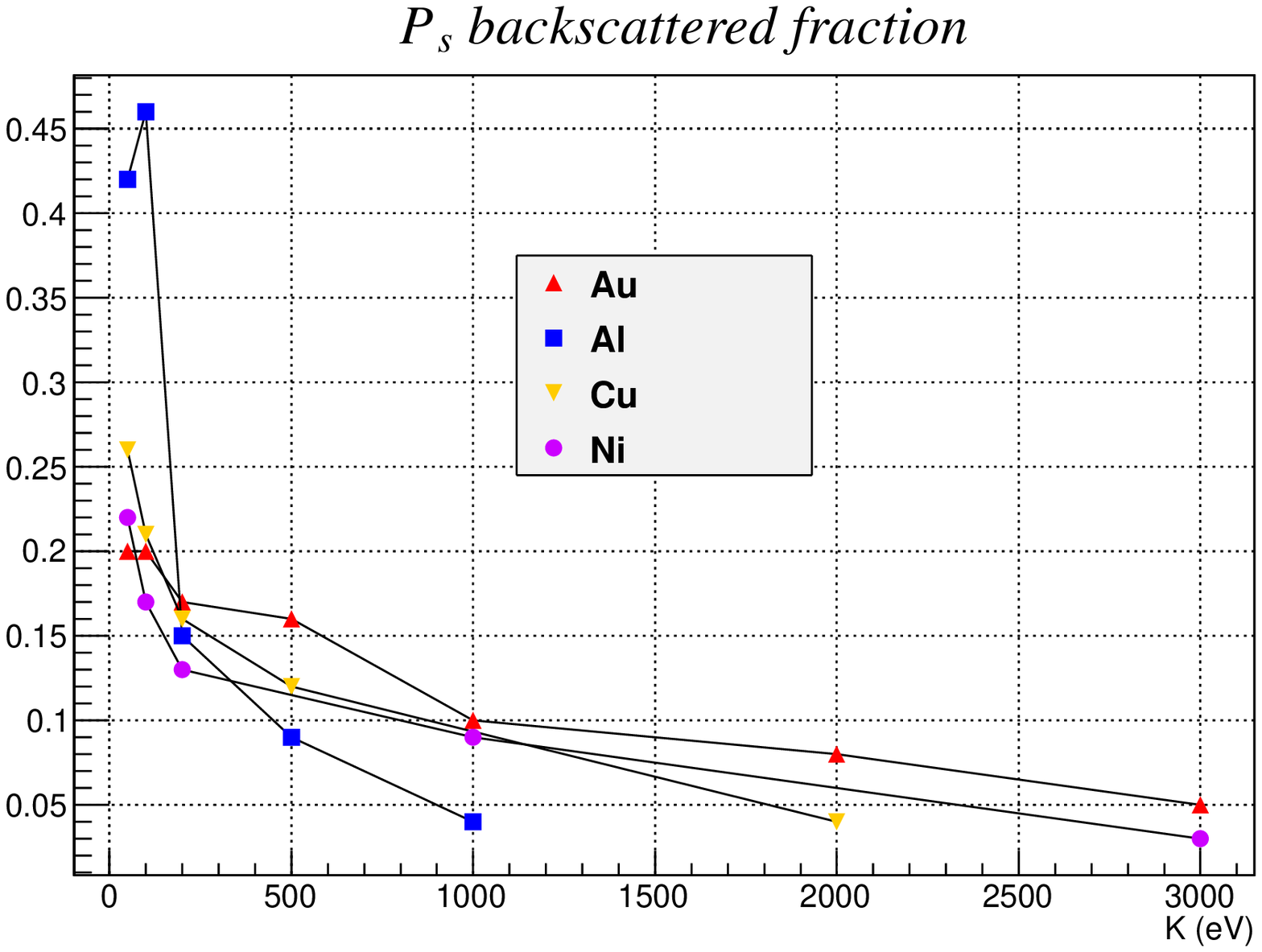} 
  \caption{}
  \label{my_fig1}
\end{figure}

\vspace{2mm}
\par
The authors have interpreted these data by comparing them with elastic backscattering of electrons, and concluded that the increase of the fraction of backscattered $(P_{s})$, resulted from the overlapping of the tail of the elastic backscattering cross section with the $(P_{s})$ formation cross section. This picture is not objected here but a complement is added. The elastic scattering is re-interpreted as diffraction, and the crossing of the curves is explained by the availabilty of Surface Plasmons Polaritons (SPP) at, or slightly above, 6.8 eV. 
\vspace{2mm}
\par

Measurements performed with ionic crystal surfaces, LiF and NaF, have shown a large backscattering of 25 eV $(e^{+})$ in the first order Bragg reflexion \cite{backscattered9}. Similar messurements have shown $(e^{+})$ Bragg reflexion below 10 eV on Crystals of Al and Cu \cite{backscattered2}. Based on these mesasurements, the $(P_{s})$ backscattering from incoming $(e^{+})$ is modeled as a 2 steps process :
\vspace{2mm}
\par

\begin{enumerate}
\item the $(e^{+})$ diffraction by the  first plane of nuclei in the metal,
\item the interaction $(e^{+})$ with  Surface Plasmons Polaritons to produce $(P_{s})$.
\end{enumerate}

\vspace{2mm}
\par

In the 2nd step, the $(P_{s})$ atom is created by the following reaction, where $(e^{-*})$ stands for electron in the metal.
\begin{equation} \label{eq:3}
\begin{aligned}
& e^{+} + e^{-*}  \rightarrow   P_{s} + h\nu\,(6.8\,eV)
\end{aligned}
\end{equation} 

\begin{figure}[htb]
  \centering
  \includegraphics[trim={8.0cm 6.0cm 0.0cm 6.0cm},clip,scale=0.8]{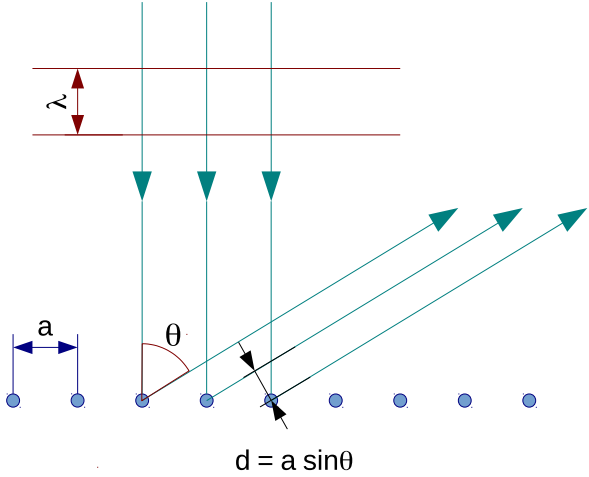} 
  \caption{Diffraction of the $(e^{+})$ by the first plan of nuclei.}
  \label{my_fig2}
\end{figure}

The density of positive charge in the first plane of atoms'nuclei prevents the positrons penetration and at low energies below $\sim200$ eV, the positrons are diffracted by a single plane. The normal incident diffraction notations are illustrated in figure (\ref{my_fig2}).

The constructive interference condition at normal incidence is then :

\vspace{2mm}
\par

\begin{equation} \label{eq:4}
\begin{aligned}
& d = a. \sin(\theta) = n. \lambda
\end{aligned}
\end{equation} 

and : 
\begin{equation} \label{eq:5}
\begin{aligned}
& \lambda = \frac{h}{\sqrt{2.m.e.V}} \approx \sqrt{\frac{150}{V}} \quad \text{\AA}
\end{aligned}
\end{equation} 

\vspace{2mm}
\par

We can now compute the angles at which the $(e^{+})$ are backscattered before capturing the $(e^{-*})$ to form $(P_{s})$ for $(n = 1)$ and  $(n = 2)$. The 4 crystals have a fcc cell, 3 are cut along the (100) plane and Al is cut along the (111) plane. Hence with a single diffracting plane : $(a = b.\sqrt{2})$ where $(b)$ is the lattice size. The results, and the corresponding lattice parameters $(b)$ are shown in table (\ref{table_2}) and table (\ref{table_3}).  

\vspace{2mm}
\par

At $100\,eV$ and above, the diffracted angle is smaller than the mask and therefore the modes $(n=0,1)$ backscattered $(P_{s})$ atoms are within the detector acceptance, but at these energies, the reaction (\ref{eq:3}) cross section decreases very fast, leading to the decrease of the backscattered $(P_{s})$ atoms.

\vspace{2mm}
\par

At $50\,eV$ only the mode $(n=0)$ is within the 30 degres mask acceptance and there are 2 allowed modes.
\vspace{2mm}
\par

\begin{table}[h]
\centering
\caption{Angle of $(n = 1 )$ diffracted $e^{+}$ with and incident energy from $(50 \, eV)$ to $(200 \, eV)$}
\label{table_2}
\begin{tabular}{|l|l|c|c|c|}
\hline
  metal & $b$ (\AA) &  $\theta \deg \; (50 \, eV)$  & $\theta \deg \; (100 \, eV)$ & $\theta \deg \; (200 \, eV)$\\
\hline
Al    & 4.046  & 37.358 & 25.346 & 17.620 \\
Cu   & 3.597  & 42.920 & 28.785 & 19.907 \\
Au   & 4.065  & 37.055 & 25.220 & 17.535 \\
Ni    & 3.499  & 44.431 & 29.670 & 20.489 \\
\hline
\end{tabular}
\end{table}

\begin{table}[h]
\centering
\caption{Angle of $(n = 2 )$ diffracted $e^{+}$ with and incident energy from $(50 \, eV)$ to $(200 \, eV)$}
\label{table_3}
\begin{tabular}{|l|l|c|c|c|}
\hline
  metal & $b$ (\AA) &  $\theta \deg \; (50 \, eV)$  & $\theta \deg \; (100 \, eV)$ & $\theta \deg \; (200 \, eV)$\\
\hline
Al    & 4.046  & - & 58.890 & 37.258 \\
Cu   & 3.597  & - & 74.377 & 42.920 \\
Au   & 4.065  & - & 58.449 & 37.055 \\
Ni    & 3.499  & - & 81.901 & 44.431 \\
\hline
\end{tabular}
\end{table}
\vspace{2mm}
\par

The lower fraction of  backsacttered $(P_{s})$ for Au, Cu, Ni relative to Al at $50\,eV$ cannot be explained by the diffraction angle, but by Surface Plasmons Polaritons, as discussed in the next section.

\vspace{2mm}
\par

With only 1 mode within the mask acceptance the fraction of backscattered $(P_{s})$ is $(\sim40\%)$ for Aluminium at $(50 \,eV)$. Hence the total fraction of  $(e^{+})$ backscattered as $(P_{s})$ at that energy is expected to be $(70\sim80\%)$.

\vspace{2mm}
\par
Below 10 eV, the adsorption of both $(e^{+})$ and $(P_{s})$ by surface states shall be taken into account \cite{backscattered8}. So the operational $(P_{s})$ beam energy range is between $\sim$10 \,eV and 100 eV.

\vspace{2mm}
\par

The divergence opening angle of the $(P_{s})$ beam can be infered from the backscattering of electrons. The Low Energy Electron Diffraction (LEED) method is a common method to investigate the surface of crystals, and the divergence of a diffracted mode is less than 10 mrd. In the case of electrons, a main contributor to the divergence comes from the diffraction of multiple ions planes, as the electrons penetrate deeper. In the case of low energy positrons, the repulsion of the first positive nuclei plane reduces this effect.

\subsection{$P_{s}$ creation and surface states}

The reaction (\ref{eq:6}) below is a radiative recombination in vacuum. The resonant energy of the emitted photon is 6.8 eV, and its small width, explain the small cross section. Conversely in a 3 body recombination like the reaction (\ref{eq:7}), the extra $( e^{-})$ provides the resonant condition by adjusting its kinetic energy. That is made possible because a free electron in vacuum can take any kinetic energy. 

\begin{equation} \label{eq:6}
\begin{aligned}
& e^{+} + e^{-}  \rightarrow   P_{s} + h\nu\,(6.8\,eV)
\end{aligned}
\end{equation} 
\begin{equation} \label{eq:7}
\begin{aligned}
& e^{+} + e^{-}  + e^{-}  \rightarrow   P_{s} + e^{-} 
\end{aligned}
\end{equation} 

For the reaction (\ref{eq:3}), the radiated 6.8 eV photon is absorbed by the metal surface states, and it shall correspond to a permitted transition. So without a resonance in the spectrum of the electrons collective surface states, a fraction of the $(e^{+} , e^{-})$ radiative recombinations does not take place. The explanation proposed for the high rate of $(P_{s})$, for Al, with $(e^{+})$ at 50 eV kinetic energy, is the availability of Surface Plasmon Polariton (SPP) resonance at 6.8 eV, or nearby with a large width, which enhances the $(P_{s})$ production cross section.

\vspace{2mm}
\par

The SPP of most metals are below the required 6.8 eV, and they cannot absorb this energy : the situation is similar to radiative capture in vacuum with a 2 body reaction. But the SPP resonant energy for Al is exceptionally high at 10.5 eV and is few eV wide, allowing the SPP resonance to overlap with the $(P_{s})$ potential well at 6.8 eV. Conversely the SPP resonant energies of (Au, Cu, Ni) are at (1.58 eV, 2.16 eV, 1.65 eV). One can infer then, that the reason for the larger fraction of $(P_{s})$ atoms emitted with a 50 eV beam on Al is the availability of a Surface Plasmon Polariton resonance.   

\vspace{2mm}
\par

The difference between the thermalized regime for incident $(e^{+})$ on Al(100) above 1 keV, and backscattered regime below 100 eV, may explain the observed inconsistency of the experimental data for incident energy at 200 eV and 1.5 keV, with both the model of positrons in the image correlation potential well, and with the model of weakly bound $(P_{s})$ atoms\cite{backscattered4}. At  1.5 keV the positrons are thermalized, while at 200 eV there is a mixing of thermalized and backscattered positrons. In this measurement\cite{backscattered4}, the divergency of the diffracted spot was found to be 7.1 mrd which is consistent with $\sim$10 mrd upper limit given above.

\vspace{2mm}
\par

An experiment was performed with a non-metal : graphite\cite{backscattered5}. It was observed that the thermalized positrons form $(P_{s})$ atoms, and these atoms are emitted within 4 mrd, while the emission intensity  grows with temperature. This last feature is a signature of the creation of $(P_{s})$ atoms from positrons which are first thermalized, then trapped at the surface. The incident $(e^{+})$ kinetic energy was 750 eV which is consistent with thermalized positrons. But the shape of the angular emission could not be explained by a direct electron capture. The model, proposed by the authors, states that the emission resulted form interactions with phonons. The comparison of the model with the data \cite{backscattered5} shows a good agreement.

\vspace{2mm}
\par

So in a similar way as the low energy $(e^{+})$, backscattered from the first layer of a metal atoms, form $(P_{s})$ atoms assisted by SPP, the high energy $(e^{+})$, thermalized inside graphite, form $(P_{s})$ atoms assisted by phonons. Increasing the phonons density with temperature, increases the emission intensity. 

\vspace{2mm}
\par

This analysis suggests two ways to improve the $(P_{s})$ beam intensity. The first one is to choose a metal with a SPP resonance at 6.8 eV. Magnesium has a SPP resonance at $\sim7$ eV and should be a suitable crystal target. The second one is to use stimulated emission. Radiative  ion electron recombination in vacuum can be enhanced by stimulated emission of the radiated photon, using a laser. In a similar fashion, reaction (\ref{eq:3}) could, in principle, be enhanced in Al and Mg by generating SPP while the $(e{+})$ beam hit the target. Such SPP could be created by an electron beam of 7 eV for Mg, or 10.5 eV for Al. SPP could also be created by photons\cite{SPP-Photon}.

\vspace{2mm}
\par

Furthermore, the launch of polarized Surface Plasmons Polaritons was demonstrated\cite{Plasmon-polarization}. Hence with stimulated emission of polarized Surface Plasmons Polaritons, if the incident $(e^{+})$ beam is also made polarized\cite{positron-polarized}, it should be possible to produce a polarized ortho-positronium beam.

\subsection{The $(P_{s})$ beam configuration.}

The diffraction configuration discussed in the previous sections, with an incident beam at $90$ degres, is the one used both by the article's data and by the LEED technique for which a large number of experimental data are available. But one could also use the Bragg angle configuration with the same metals, Al and Mg. The analysis is the same and the conclusions hold.  

\vspace{2mm}
\par

\begin{figure}[htb]
  \centering
  \includegraphics[trim={4cm 10.5cm 4cm 8cm},clip,scale=0.5]{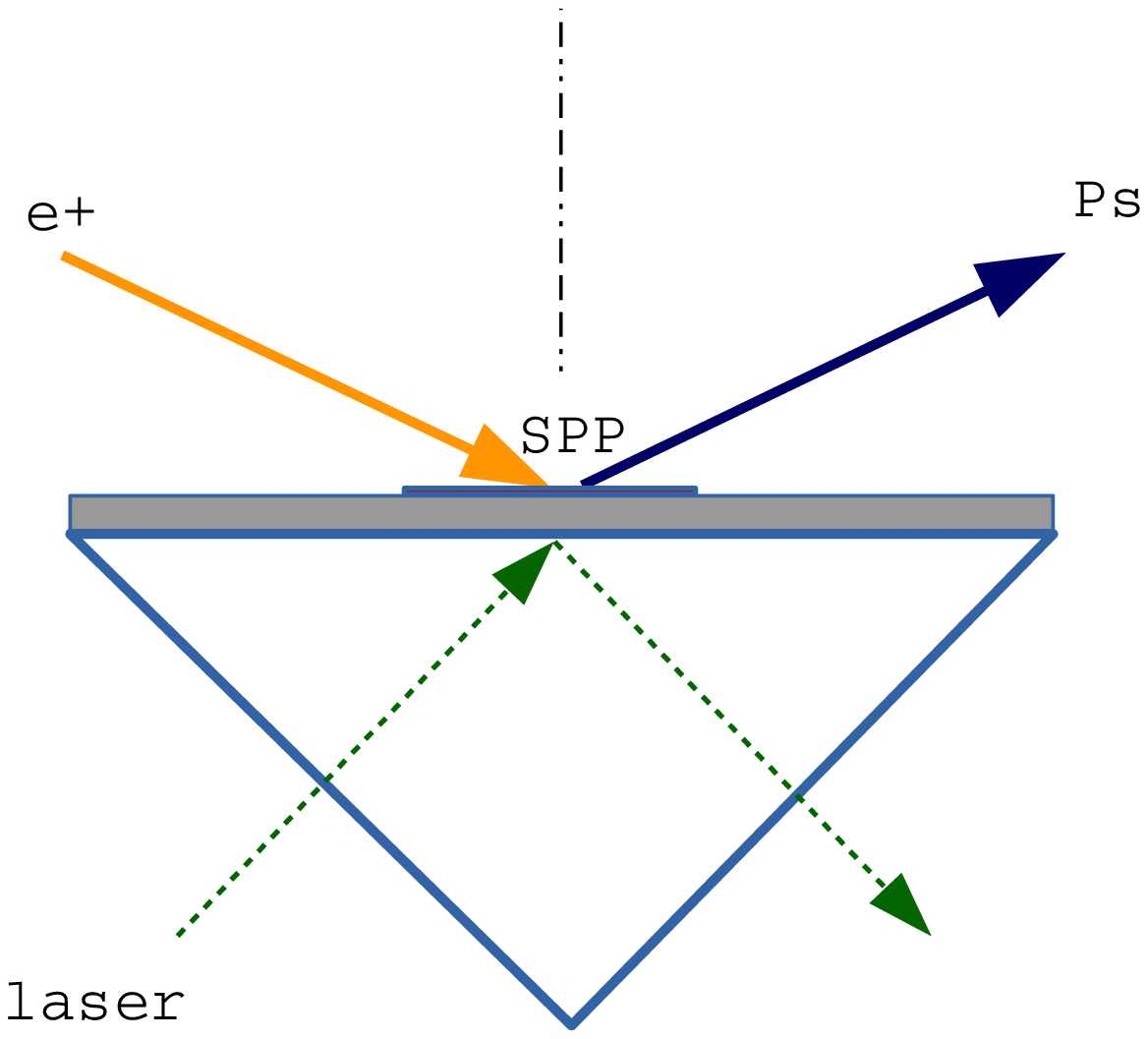} 
  \caption{Layout for an SPP stimulated emission of a $(P_{s})$ atom beam.}
  \label{my_fig3}
\end{figure}

The apparatus proposed to create a $(P_{s})$ beam, is made of a crystal with a SPP resonance overlapping 6.8 eV and with an energy tunable $(e^{+})$ beam. A possible configuration is illutrated in figure \ref{my_fig3}. This beam is made incident at a selected Bragg angle. With an incident energy below 100 eV, the $(e^{+})$ do not penetrate deep in the crystal, and only the first two planes have to be taken into account for the Bragg diffraction. These effect is reflected in the very high backscattering efficiency observed in the measurement at 90 degrees incident beam discussed in the previous section. With Aluminium, the expected efficiency of emitting a $(P_{s})$ atom in a single selected mode at Bragg reflexion is the same as the estimated total fraction of  $(e^{+})$ backscattered as $(P_{s})$ atoms, that is $70\%$. The $(e^{+})$ which are reflected without capturing an electron are few. Above 10 eV, the measurements on Al give a $\sim2\%$ reflection of $(e^{+})$ exiting the crystal : most of the incident $(e^{+})$ are reflected as $(P_{s})$ atoms\cite{backscattered}\cite{backscattered2}.

\vspace{2mm}
\par

This estimated efficiency is a lower limit, since no stimulated emission of SPP and no polarization of both the SPP and the $(e^{+})$ were implemented. Hence with a single mode emission at a selected Bragg angle, it is expected that $70\sim80\%$ of the $(e^{+})$ beam is transformed into a $(P_{s})$ beam. And a higher efficiency is expected with Magnesium and/or with stimulated SPP emission.

\vspace{2mm}
\par

Unlike for $(e^{-})$, there is no Fermi sea of positrons in the metal, and the absolute value of the work function is very small, hence the $(e^{+})$ are diffracted with their incident kinetic energy. The energy variation is the 6.8 eV emission for the creation of $(P_{s})$ corrected by the small work function.

\vspace{2mm}
\par

\section{$(\bar{H}^{+})$ production with a tunable $P_{s}$ beam}

The GBAR experiment will be taken as reference for the discussion, but the same arguments are valid for similar experiments based on $(\bar{H}^{+})$.

\subsection{Kinematics}
The kinematic energy of the reaction (\ref{eq:1}) is better understood in the reference frame of the $(p^{-})$ at rest. An incoming $(P_{s})$ atom requires at least 6.8 eV to break the link between the positron and the electron. This energy is the depth of the potential of the  $(P_{s})$ atom in the ground state. This energy is provided by the kinetic energy of the incoming $(P_{s})$ atom. So the cross section for producing  $(\bar{H})$ starts at 6.8 eV. At larger kinetic energy, above $\sim$ 40 eV, the positron $(e^{+})$, once released by the breaking of  $(P_{s})$, carries to much kinetic energy to fall into the 13.6 eV potential of $(\bar{H})$. Hence the cross section is high between 10 eV and 40 eV, and neglectable at others $(P_{s})$ beam kinetic energies.   

\vspace{2mm}
\par

Once transposed in the reference frame of the $(P_{s})$ atom at rest, with a $(p^{-})$ beam, the kinetic energy of the antiproton for the maximum cross section is around 10$\sim$15 keV due to the fact that its mass is $(\sim 1000 \times)$ larger than the one of the $(P_{s})$ atom.

\vspace{2mm}
\par

When a light object in a beam, $(P_{s})$, collides with an heavy object at rest, in this case $(p^{-})$, the heavy object does not recoil, and all the kinetic energy of the incoming light object is converted to break the $(P_{s})$ atom. Conversely when an heavy object collides with a light one at rest, the light object 's recoil absorbs most of the incoming kinetic energy which is no more available for the reaction. That is why, a 10 eV $(p^{-})$ beam colliding with a $(P_{s})$ target at rest will not produce the reaction (\ref{eq:1}).

\vspace{2mm}
\par

 The table (\ref{table_0_1}) gives for each bin of $(P_{s})$ kinetic energy in the frame of $(p^{-})$ at rest, the corresponding kinetic energy of the $(p^{-})$ in the frame of the $(P_{s})$ atom at rest.

\vspace{2mm}
\par

\begin{table}[h]
\centering
\caption{$(P_{s})$ kinetic energy with $(p^{-})$ at rest \\ and equivalent $(p^{-})$ kinetic energy for $(P_{s})$ at rest (eV)}
\label{table_0_1}
\begin{tabular}{|l|l||l|l||l|l|}
\hline
 $K_{P_{s}}$  & $K_{p^{-}}$ & $K_{P_{s}}$  & $K_{p^{-}}$  & $K_{P_{s}}$  & $K_{p^{-}}$ \\
\hline
  1 &   918.083 & 11  &  10098.9 & 21 & 19279.7 \\
  2 &  1836.17  & 12  &  11017    &  22 &  20197.8 \\
  3 &   2754.25 & 13  &  11935.1 & 23  &  21115.9  \\
  4 &   3672.33 &14  &  12853.2  & 24  &  22034 \\
  5 &  4590.41  &15  &   13771.2 & 25  &  22952.1\\
  6 &  5508.5    &16  & 14689.3   & 26  &   23870.1\\
  7 &   6426.58 & 17 &  15607.4  & 27 & 24788.2  \\
  8 & 7344.66   &18  & 16525.5  & 28  &  25706.3  \\
  9 &   8262.74 & 19 & 17443.6  & 29  &   26624.4 \\
10 & 9180.83   & 20 & 18361.7  & 30  &  27542.5 \\
\hline 
\end{tabular}
\end{table}


\vspace{2mm}
\par

For the next step, reaction (\ref{eq:2}), to get the ion, the electromagnetic potential of the second positron $(e^{+})$ is at the depth of 0.7 eV, and the depth of the (n=3) excited state of the $(P_{s})$ atom is 0.75 eV : the very near depth of the two potentials allows for a very high cross section, provided that the relative speed of $(P_{s})$ and $(\bar{H})$ is small. The contradiction is then the requirement to have a 15 keV  $(p^{-})$ beam which, once transformed into $(\bar{H})$, shall become a nearly 0 eV beam.

\vspace{2mm}
\par
The solution proposed here is different : use a tunable 10 eV $(P_{s})$ beam on $(p^{-})$ at rest, to optimize both the $(P_{s}(n=1))$ interaction of reaction (\ref{eq:1}), and the $(P_{s}(n=3))$ interaction for the reaction (\ref{eq:2}). This solution hence relies on the availability of a $(P_{s})$ beam with a tunable kinetic energy at $\sim10$ eV.

\vspace{2mm}
\par

Without a tunable $(P_{s})$ beam, GBAR has to find a compromise between the collision kinetic energies suitable for both the reactions (\ref{eq:1}) and (\ref{eq:2}). Since GBAR is a $(p^{-})$ beam on a $(P_{s})$ target experiment, the compromise is on the $(p^{-})$ beam energy which is set at 6 keV. 

\vspace{2mm}
\par

Conversely here the first reaction is a $(P_{s})$ beam colliding on a slow moving $(p^{-})$ target, and the second one is  a $(P_{s})$ beam in the  $(3^{3}P)$ state crossed by an $(\bar{H})$ beam. 
\vspace{2mm}
\par

The initial $(p^{-})$ target will be in the [20 eV , 100 eV] window. The kinetic energy has two functions :
\begin{enumerate}
\item to extract the $(p^{-})$ cloud from the trap where it is cooled,
\item to package the cloud into a beam 2 cm long, with a diameter below 1 mm, beam which defines the line of flight. 
\end{enumerate}

From the point of view of the reactions (\ref{eq:1}) and (\ref{eq:2}), this energy range is so low that the target is \enquote{at rest}. While at GBAR the $(P_{s})$ cloud is enclosed inside a square tube 1 x 1 mm$^{2}$ and 2 cm long, here the $(P_{s})$ will be a beam that is first used in a collision mode to produce $(\bar{H})$, and then slown down, while being set in the $3^{3}P$ state, so that the produced $(\bar{H})$ is slightly faster and crosses the excited $(P_{s})$ cloud. The same $(P_{s})$ beam is used for both reactions, but the preparation of the $(P_{s})$ in the excited $(3^{3}P)$ state is used to slow down the beam at a speed slightly lower than the $(p^{-} , \bar{H})$ beam.

\subsection{The $(p^{-})$ target and the steps sequence}

The $(p^{-})$ target is easier to handle if it is not at rest. And in order to get the largest cross sections for the reactions (\ref{eq:1}) and (\ref{eq:2}), the $(P_{s})$ beam shall hit the target at $\sim$ 10 eV, and then be slown down to the speed of $(\bar{H})$.
\vspace{2mm}
\par

The following configuration is proposed:

\begin{enumerate}
\item start with a $(p^{-})$ target at low energy, between 20 and 100 eV,
\item hit the target by a faster $(P_{s})$ beam in the same direction at 10 eV for reaction (\ref{eq:1}),
\item after the $(P_{s})$ atoms are ahead of the $(p^{-} , \bar{H})$ beam, slow down a fraction of the $(P_{s})$ atoms with a $(1^{3}S-3^{3}P)$ transition (205 nm, 6.05 eV),
\item then slown down a fraction of the $(P_{s})$ atoms which are in the $(3^{3}P)$ excited state, with a $(3^{3}P-2^{3}S-3^{3}P)$ stimulated emission + excitation  (1320 nm , 2 x 0.94 eV)),
\item repeat step 4 once,
\item the $(\bar{H})$ being slightly faster than the excited $(P_{s})$ beam, the collision is at low relative speed to produce the ion $(\bar{H}^{+})$.
\end{enumerate}

\vspace{2mm}
\par
The steps 1 and 2 do not involve any laser interaction and are the only steps required to produce an $(\bar{H})$ beam.

\vspace{2mm}
\par

The steps 3 and 6 are required to go from $(\bar{H})$ to $(\bar{H}^{+})$.
\vspace{2mm}
\par
The steps (4,5) are optional.

\vspace{2mm}
\par
The transitions to/from excited states do not involve any cooling sequence : the $(1^{3}S-3^{3}P)$ transition is only made once  and the  $(3^{3}P-2^{3}S-3^{3}P)$ transition are only made twice. Furthermore the excited atoms which decays through a spontaneous emission are lost and no attempt is made to collect them with a tube like in GBAR, or with lasers in the transverse plane. 
\vspace{2mm}
\par
The goal is exclusively to shift the peak of the speed distribution down so that the speed of the $(\bar{H},p^{-})$ beam becomes larger than the one of the $(P_{s}(n=3)$). The cross section are not sensitives to a spread of the energy distribution of 1 eV and there is no need to reduce the width of the atoms speed distribution, hence no cooling. But the lasers shall have a wide line width, which covers Doppler broadening without sweeping the wavelength.

\vspace{2mm}
\par
After the $(P_{s})$ beam has crossed the region where the $(1^{3}S-3^{3}P)$ laser beam is focussed, the $(P_{s})$ atoms can be in one the 3 following states:
\begin{itemize}
\item $(P_{s})$ has not absorbed a photon and continues in the beam unaffected,
\item $(P_{s})$ has absorbed a photon but has not yet emitted a photon by spontaneous decay, hence it is still in the beam,
\item $(P_{s})$ has absorbed a photon and has decayed emitting the photon in any direction, the $(P_{s})$ atom is lost.
\end{itemize}

\vspace{2mm}
\par
Starting from the $15\%$ efficiency for the $(1^{3}S-3^{3}P)$ transition from \cite{AEGIS_LASER} and with several lasers :
\begin{itemize}
\item after the 1st focussed region $15\%$ of $(P_{s})$ is either in the $3P$ state in the beam or lost,
\item after a 2nd focussed region $12.75 \%$ (0.15x0.85) of $(P_{s})$ is either added to the fraction of $3P$ atoms in the beam or lost,
\item after a 3rd focussed region $10.84 \%$ of $(P_{s})$ is either either added to the fraction of $3P$ atoms in the beam or lost,
\item etc.
\end{itemize}

\vspace{2mm}
\par
Giving a detailed description of a beam on target experiment, with a $(P_{s})$ beam hitting $(p^{-})$ target at rest, is beyond the scope of this article which is to discuss the possibility of that configuration with a $(P_{s})$ beam and to give an order of magnitude of the production rates. Within that limited precision, it is assumed that using several lasers for the $1S-3P$ transition, the fraction of atoms which absorb a photon is large enough to compensate for the losses. With 4 lasers, $15 + 12.75 + 10.84 + 5.79 = 44.38\%$ of the atoms makes the $1S-3P$ transition. In the following, to estimate the rates, the approximation is made that the initial number of excited Ps(n=3) atoms can be taken as reference, $15\%$, and that all the losses are merged into this number.

\vspace{2mm}
\par
The rates are estimated with all the 6 steps and the  $(p^{-})$  beam energy peak is set at 80 eV.
\vspace{2mm}
\par

The 2 transitions $(3^{3}P-2^{3}S-3^{3}P)$ will be refered to as a \enquote{Slow Down Station} (SDS). Other similar sequences can be done via n=2 and/or n=4 states to reach an excited $n=3$ state with  $(P_{s})$ final kinetic energy below 1 eV. The $(1^{3}S-3^{3}P)$ is selected here because it has been proven experimentally\cite{AEGIS_LASER}, and because the full SDS sequence with these 3 states is relatively simple compaired to others paths. 
\vspace{2mm}
\par

Once the steps 3,4,5 have taken place, the peak of the $(P_{s})$ beam kinetic energy distribution falls down to $10 - 6.05 - 4 \times 0.94 = 0.17 $ eV. To estimate the$(e^{+})$ and the $(p^{-})$ beam kinetic energy distributions in the transport lines require a simulation and is difficult to compute. But by tuning the $(e^{+})$ and the $(p^{-})$ initial beams energies, one can reach experimentally the peak production rates : the cross section of reaction (\ref{eq:2}) can be optimized while keeping the reaction (\ref{eq:1}) near its maximum.

\vspace{2mm}
\par

One of the advantage of using very low energy $(p^{-})$ beam is that relative to the $(p^{-})$ mass, this energy is so small that even if poorly defined, it does not affect the cross section. Conversely in the $(p^{-})$ beam on $(P_{s})$ target at rest configuration, the $(p^{-})$ beam spread of kinetic energy of $\sim$1.5 keV degrades either reaction (\ref{eq:1}) or reaction (\ref{eq:2}). 


\vspace{2mm}
\par

Note that to produce $(P_{s})$ at 10 eV, the incident $(e^{+})$ beam shall be at 10 + 6.8 = 16.8 eV because  6.8 eV will be absorbed by the emission of a Surface Plasmon Polariton.

\subsection{The experimental layout}

\begin{figure}[htb]
  \centering
  \includegraphics[trim={2cm 10cm 4cm 9cm},clip,scale=0.5]{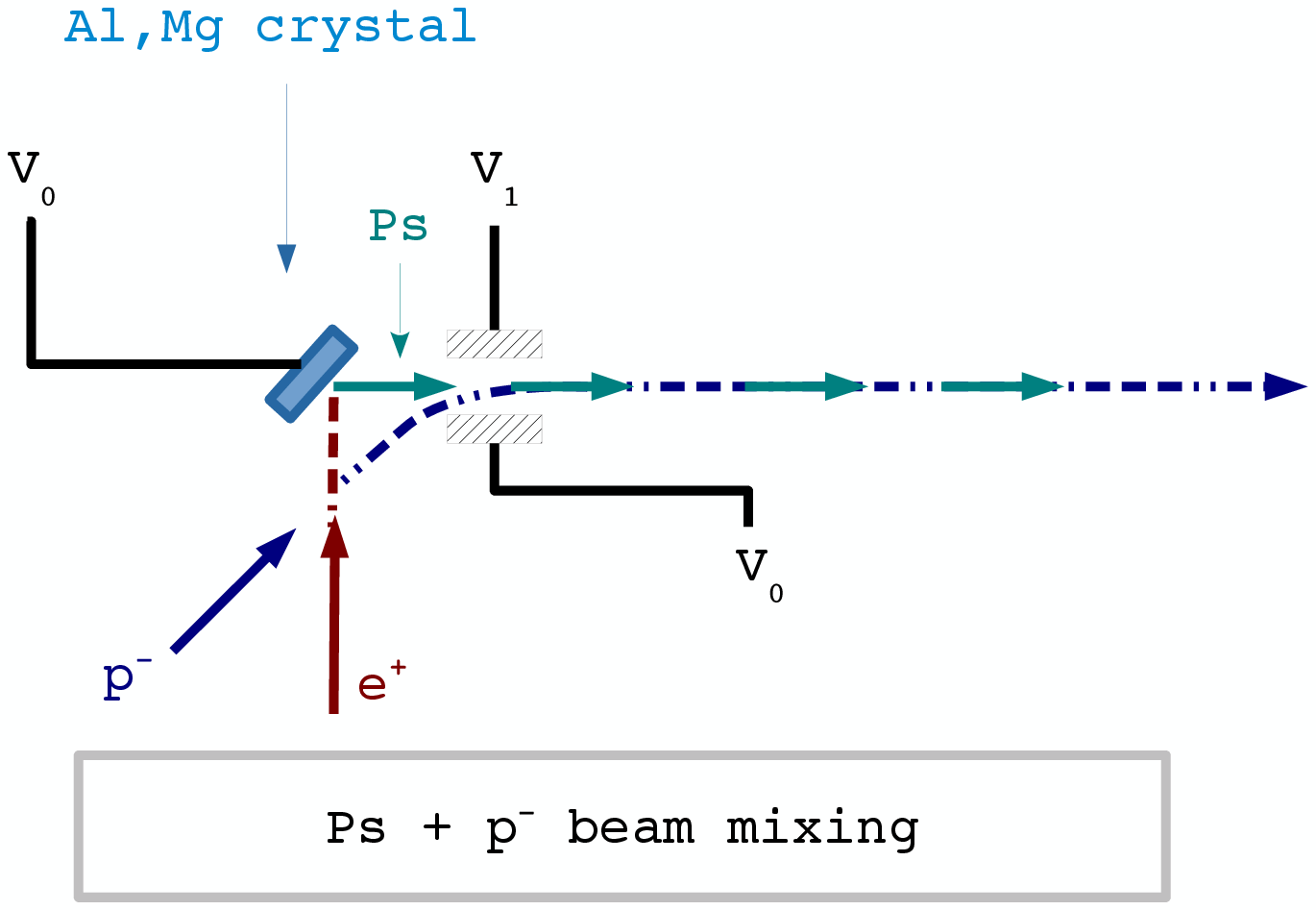} 
  \caption{}
  \label{my_fig4}
\end{figure}

The experimental layout to implement this configuration is drawn in figure \ref{my_fig4}. With a pair of electrostatic deflectors, the $(p^{-})$ beam can be aligned with the $(P_{s})$ beam within 1 cm of flight length.

\vspace{2mm}
\par

Starting with a $(P_{s})$ beam with a diameter of 100 $\mu$m, and with a divergence of 10 mrd (10 mrd is conservative, smaller divergences are obtained with LEED), the $(P_{s})$ beam diameter grows 200 $\mu$m every 1 cm. Distances will be counted from the metal crystal $(P_{s})$ source. With a separation of 1 cm betwen the metal crystal, and a 1 cm long electrostatic deflector, the reactions (\ref{eq:1}) and (\ref{eq:2}) start at 2 cm with a $(P_{s})$ beam diameter of 500 $\mu$m. The beam diameters at several distances from the crystal metal $(P_{s})$ source are given in table (\ref{table_5})

\vspace{2mm}
\par

\begin{table}[h]
\centering
\caption{$P_{s}$ beam diameter as a function of the distance from the crystal.}
\label{table_5}
\begin{tabular}{|l|l|l|l|l|l|l|}
\hline
 x (cm)  & 1 & 2 & 3 & 4 & 5 & 6  \\
 $\phi$ ($\mu$m) & 300 & 500 & 700 & 900 & 1000 & 1200 \\
\hline
\end{tabular}
\end{table}

\vspace{2mm}
\par
The following hypothesis are made on the $(e^{+})$ and $(p^{-})$ beams. 
\vspace{2mm}
\par
Small positron microprobe beams have been developped with remoderation techniques but the positrons final kinetic energy was in the keV range and not at 16.8 eV. In a positron microprobe beam, after being accelerated, the positrons are focussed on a thin moderator foil ($100\sim200$ nm Ni or W) where they are thermalysed before being re-emitted on the other side. The beam size after 2 remoderation steps can be as small as 10 $\mu$m and the energy spread as small as the thermal energy $\sim10$ meV \cite{Brightness}. In the scanning positron microprobe these positrons are accelerated to several keV before hitting the sample. For creating the $(P_{s})$ beam, the positrons shall be accelerated to 16.8 eV. Hence an electrostatic focussing/acceleration device few mm long has to be designed. This procedure is not proven. The hypothesis is made that this step is feasible and that one can produce a 100 $\mu$m diameter $(e^{+})$ beam with a divergence below 10 mrd at 16.8 eV. The remoderation provides also the beam length compression. A 66 cm long $(e^{+})$ beam at 10 keV will be absorbed by the moderator in less than 11.4 ns, so the reaccelerated $(P_{s})$ beam at 16.8 eV will be $\sim 2$ cm long. 

\vspace{2mm}
\par
With a rotating electric field method, 5 $10^5$ antiprotons have been cooled down to 0.3 eV, packaged into a beam with a radius of 0.25 mm and extracted from a trap with a kinetic energy of 50 eV \cite{antiprotoncompression}. This method was not tested with $10^7\sim10^8$ antiprotons. It is assumed that these numbers of antiprotons can be acheived. Furthermore the extracted beam would have to be accelerated to 80 eV and compressed into a 2 cm long beam. The compression increases the beam divergence, but the beam expansion will be neglected on the few cm propagation region where the $(P_{s},p^{-})$ interaction takes place. A dedicated electostatic device  would have to be designed. This procedure is not proven.

\vspace{2mm}
\par

\vspace{2mm}
\par
Once the $(P_{s})$ beam has hit the $(p^{-})$ beam and produced a $(p^{-} , \bar{H})$ beam, it is slown down to $\sim$0.06 eV with the lasers. This step is implemented when the two beams overlap, that is when coming from the back, the $(P_{s})$ beam has fully covered and hit the $(p^{-})$ beam.

\vspace{2mm}
\par

Since the speed of $(P_{s})$ at 0.06 eV  is $\sim$1.02 $10^{4}\,m.s^{-1}$ and the speed of a $(p^{-} , \bar{H})$ with 80 eV kinetic energy is 1.02 $10^{5}\,m.s^{-1}$, $(\bar{H})$ moves $\sim10\times$ faster than $(P_{s})$. So if both beam are 2 cm long, it will take 200 ns for the $(\bar{H})$ to cross the $(P_{s})$ beam. The lifetime of $(P_{s}(n=3))$ being 11 ns, the region of laser excitation shall extend over $\sim3$ cm and at least 4 lasers $(1^{3}S-3^{3}P)$ shall be used.

\vspace{2mm}
\par

Some $(\bar{H})$ will not capture an $(e^{+})$, and will continue the flight as well as the $(P_{s})$ atoms which are not slowndown or lost by a spontaneous decay to the ground state. The non excited $(P_{s})$ atoms will decay spontaneousely in 142 ns. With a speed of $\sim2\,10^{6}\,m.s^{-1}$ they decay in $\sim$30 cm. After $\sim1$ m, the only particles left in the beam are $(p^{-} , \bar{H}, \bar{H}^{+})$.

\subsection{$(\bar{H}^{+})$ in numbers with CUSP and GBAR using a $(P_{s})$ beam.}

CUSP was designed to measure the n=1 hyperfine transition of $(\bar{H})$, and it uses a radioactive $(e^{+})$ source much less intense than the linac source used by GBAR. Nevertheless, the CUSP source would be just sufficient to perform a free fall experiment, and  CUSP parameters give the threshold limit at which this method can be used. This section will compair the number of $(\bar{H})$ and $(\bar{H}^{+})$ that can be produced by both experiments. The layout modification for CUSP would be to replace the $(\bar{H})$ apparatus by the one proprosed here and to keep the hyperfine transition measurement apparatus.

\vspace{2mm}
\par

CUSP can store 2 10$^7$ $(e^{+})$ in its trap, and 10$^7$ $(p^{-})$.

\vspace{2mm}
\par

We will asume that the $(p^{-})$ beam is 2 cm long and has a diameter of $\sim$0.85 mm. Both $(p^{-})$ and $(e^{+})$ are supposed to be cooled down. The energy spread of the $(p^{-})$ will have to be measured experimentally. As long as the $(p^{-})$ have a small kinetic energy, less than $\sim100$ eV, they are approximatively at rest realtive to the $(P_{s})$ beam. The design of electrostatic devices for focussing the $(p^{-})$ and  $(e^{+})$ beams is beyond the scope of this article. The purpose of this section is to give an order of magnitude of the production rates.

\vspace{2mm}
\par

With these numbers and using a conservative Al crystal without stimulated SPP emission, $\sim70\%$ of the $(e^{+})$  delivered in a 2 cm long beam at 16.8 eV will produce the $(P_{s})$ beam at 10 eV. Hence the 10 eV $(P_{s})$ beam will carry 1.32 10$^{7}$ atoms. The volume is given by the average diameter of the $(P_{s})$ beam between L = 3 cm and L = 5 cm. Its inverse, 1/volume, is 176 which gives the  beam density for reaction (\ref{eq:1}).

\begin{equation} \label{eq:8}
\begin{aligned}
&\rho = 176 \times 1.32 \, 10^{7} =  2.32 \, 10^{9} \,cm^{-3}
\end{aligned}
\end{equation}

 The $(P_{s})$, produced in the ground state, being much faster than the $(p^{-})$ beam, the later will be considered at rest for reaction (\ref{eq:1}). Around 10 eV, a $(P_{s})$ beam on $(p^{-})$ at rest has a cross section of $\sim$15 10$^{-16}$ cm$^{2}$, and the number of $(\bar{H})$ produced at each shot is then :

\begin{equation} \label{eq:9}
\begin{aligned}
&2.32 \, 10^{9} \times 10^{7} \times 15 \, 10^{-16} \times 2 = 69.6 \quad  (\bar{H})
\end{aligned}
\end{equation}

Following the AEGIS method\cite{AEGIS_LASER} with a single photon excitation at n=3, 2 SDS and with the approximations discussed in the previous sections, the fraction of $(P_{s})$ n=3 will be $\sim$ 15$\%$. The states n=2 and  n=1 also produce $(\bar{H}^{+})$, but they will not be taken into account in order to keep the estimation conservative. The number of $(P_{s})$ in the beam in the state n=3 is then 2 10$^6$.

\vspace{2mm}
\par
In the 2 cm overlap volume the density of excited $(P_{s})$ atoms is then :  
  
\begin{equation} \label{eq:10}
\begin{aligned}
&\rho = 3.52 \, 10^{8} \,cm^{-3}
\end{aligned}
\end{equation} 

The reaction (\ref{eq:2}) cross section with $(P_{s})$ in the $3^{3}P$ state is $\sim$ 1000 10$^{-16}$ cm$^{2}$. Then, assuming that only 1/2 of $(\bar{H})$ are produced in the first of the 2 cm, and that only 1cm is available for reaction 2, the number of $(\bar{H}^{+})$ per shot produced via the $3^{3}P$ state is :

\begin{equation} \label{eq:11}
\begin{aligned}
&3.52 \, 10^{8} \times 69.6/2 \times 1000 \, 10^{-16} \times 1 = 0.0012
\end{aligned}
\end{equation} 


As an hyperfine measurement experiment, even without ELENA, CUSP could produce 69.6 $(\bar{H})$ in a timely way, within a $\sim$100 ns window. This may be to much in single shot, but CUSP could run at a faster rate with less $(e^{+})$ or $(p^{-})$ per shot. At the moment, the experiment has no timing since the $(\bar{H})$ escape randomly from the apparatus where they are produced. A small time window would reduce the background noise.    

\vspace{2mm}
\par

GBAR will be connected to ELENA from its start in 2017, and has a linac source which delivers $\sim10^{10}$ cooled down $(e^{+})$ in its trap, that is 500 times more than CUSP. With these numbers, and assuming 10$^{8}$ $(p^{-})$ per shot, GBAR would produce 5000 mores $(\bar{H})$ and $(\bar{H}^{+})$ per shot :

\begin{itemize}
\item 348000 $(\bar{H})$ per shot,
\item 6 $(\bar{H}^{+})$ per shot.
\end{itemize}  

Since ELENA is expected to deliver $4\,10^{6}$ $(p^{-})$ every 110 seconds, GBAR would have to accumulate 25 burst to get 10$^{8}$ $(p^{-})$.


\section{Gravitational interferometry}

An interferometry with $(P_{s})$ was suggested in 2002 \cite{Ps_interferometry}, using the light wave Bragg diffraction for Raman splitters and mirors, which made the exit beam intensity very weak. In 2002, the experiment was hardly feasible. Today, with more powerfull lasers and $(e^{+})$ sources, it might be within reach.

\vspace{2mm}
\par

The solution proposed here is simpler : since the $(P_{s})$ are produced by the diffraction of an incident $(e^{+})$ beam, the source is coherent and the emitted modes can be used as such to create a Mach-Zehnder atom interferometer. The two beams recombination is performed by a specular reflection on a LiF crystal for each beam. Measurements have shown that between 7 eV and 60 eV specular reflexion of $(P_{s})$ atoms has an efficiency that decrease from $30\%$ to $3\%$ \cite{backscattered10}\cite{LiF}. These values are sufficient to operate an interferometer at a kinetic energy of 10 eV.

\vspace{2mm}
\par

At that energy the $(P_{s})$ de Broglie wavelength is 2.74 $\text{\AA}$. Historically, classical crystal diffraction is the path used for neutrons and electrons interferometry : no  light wave Bragg diffraction was  used. Unlike matter atoms which masses are too large and which wave lengths are too small, the $(P_{s})$ wave length is also suitable for a solid state interferometer.

\vspace{2mm}
\par

This interferometer is similar to the one used in the Colella Overhauser Werner experiment (COW).  The differences are :

\begin{itemize}
\item the neutrons are replaced by $(P_{s})$ atoms in the ground state which electroweak decay rates are measured,
\item the beam splitters are replaced by passive diffraction gratings or partially reflecting Au or LiF mirror.
\end{itemize}

Progress in nano technology opens new possibilities for splitting and recombining a $(P_{s})$ beam : gold nano mesh, nano grating and LiF mirors dispersed on a graphene sheet.

\vspace{2mm}
\par

The developpement of nano grid for Transmission Electron Microscopy (TEM), makes available gold grid, packaged within 3-mm diameter disc of gold mesh where a layer of gold foil with a regular array of micrometer-sized holes is suspended across the square openings in the mesh \cite{backscattered6c}. These micro mesh, 500 nm thick, can be used as semi-transparent miror for the $(P_{s})$ beam.

\vspace{2mm}
\par

Nano technology allows for the fabrication of grahene gratings with slit widths and separation in the $\sim$50 nm range suitable for diffracting $(P_{s})$ atoms with a wave length of $\sim2\text{\AA}$. An alternative is nanoporous alumina, which is commercialy available with pores diameter of 20 nm and a thickness from 200 nm to 200 $\mu$m. Assuming that grating in the 5 nm range can be produced, and absorbing all the diffracted modes but 2, one could reach an angular separation of $\sim$0.1 rd and use the grating for beam splitting and recombination.

\vspace{2mm}
\par

Conversely, larger angular separation could be obtained with LiF semi-transparent mirors. The developpement of Li-ion batteries has triggered a rich field of research in the methods for producing larges, stables, carbon electrodes, modified with surface deposition of LiF nano particles. Recently\cite{backscattered6b}, it has been demonstrated that LiF nano particles with a diameter of 5-15 nm can be dispersed and attached to a graphene sheet. The nano particle have a good crystallinity with (002) plane on graphene. Being oriented they can act as dispersed mirors for a $(P_{s})$ beam. Using a graphene monolayer on a TEM grid, like for instance the Quantifoil Gold produced by Sigma-Aldrich, a semi-transparent $(P_{s})$ miror could be obtained. Assuming 0.25 reflection for 10 eV $(P_{s})$ atoms, and a LiF/graphene surface ration of 0.2, the 2 reflections in the interferometer path before the recombination would give 0.0025 efficiency. Assuming 0.05 transparency with losses due to the interactions with the graphene and with the grid, crossing twice the miror would give 0.0025 efficiency. One should add the losses in the last semi-transparent miror used for the recombination, and then, the overall the proportion of $(P_{s})$ atoms exiting the Mach-Zehnder interferometer would be $\sim10^{-3},10^{-4}$.

\vspace{2mm}
\par

\begin{figure}[htb]
  \centering
  \includegraphics[trim={2.0cm 5.0cm 0.0cm 0.0cm},clip,scale=0.4]{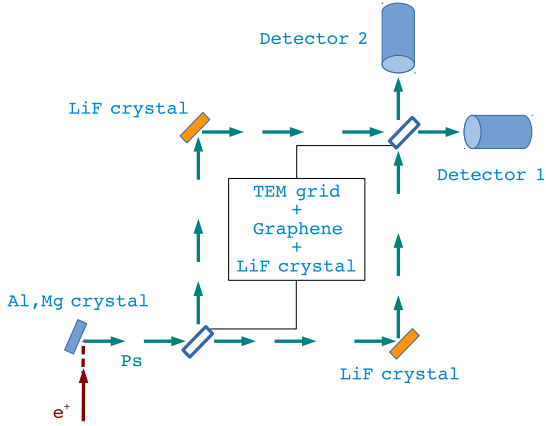} 
  \caption{Mach-Zehnder atom interferometer with a $P_{s}$ coherent beam.}
  \label{my_fig5}
\end{figure}

The layout of such interferometer is drawn in figure \ref{my_fig5}. In this layout, the mirors on each separated arm use specular reflection on a LiF crystal with 0.25 reflection coefficient.

To define the parameters, we rewrite the COW interference formula.
The COW experiment phase shift between the two arm of the interferometer is computed with classical mechanics since the atoms'speed is $\sim10^{6}m.s^{-1}$. The interferometer will be rotated around the horizontal axis. The atoms are launched horizontally and their wave is splitted in two. The atoms on the horizontal axis have a kinetic energy $E_{0}$, a wave vector $k_{0}$, and the gravitation potential origin is set at this level. Once at the upper level $z$ of the interferometer, the wave vector is $k$. The phase shift when the atoms go from the level $0$ to $z$ is the same for the two paths. Hence the phase difference at the recombination point is given by $\Delta \Phi = L . \Delta k$ with $L$ being the horizontal path length. 

\begin{equation}\label{eq:12}
\begin{aligned}
&\frac{\hbar^{2} . k^{2}}{2.m} = E_{0} - m.g.z \Rightarrow \ k^{2} = \frac{2.m.E_{0}}{h^{2}}.(2\pi)^{2}.(1 - \frac{m.g.z}{E_{0}}) \\
&k_{0}^{2} = \frac{2.m.E_{0}}{h^{2}}(2\pi)^{2}  \Rightarrow  k^{2} = k_{0}^{2}.(1 - \frac{m.g.z}{E_{0}}) \\
&m.g.z >> E_{0} \Rightarrow k \sim k_{0}.(1 - \frac{m.g.z}{2.E_{0}}) \\ 
&\Rightarrow k - k_{0} = - \frac{m.g.z.k_{0}}{2.E_{0}}
\end{aligned}
\end{equation} 

The rotation angle around the horizontal axis being $\alpha$, as the interferometer is rotated, the COW phase shift becomes :

\begin{equation} \label{eq:13}
\begin{aligned}
&\Delta \Phi = - \frac{m.g.z.L.\sin \alpha .k_{0}}{2.E_{0}} = - \frac{m.g.z.L.\sin \alpha .k_{0}}{2.( \hbar^{2} . k_{0}^{2} / 2.m  )} \Rightarrow \\ 
&\Delta \Phi = \frac{-2\pi .g.\lambda .m^{2}.z.L.\sin \alpha}{h^{2}}
\end{aligned}
\end{equation} 

The $(P_{s})$ lifetime is 142 ns, so its half lifetime is 98 ns. The corresponding distance, before 1/2 of the $(P_{s})$ are lost, is $L_{1/2}$. The product $z.L$ is the surface of the interferometer. It is set to $S = (L_{1/2} / 2)^{2}$. With these settings, the phase shift can be written $\Delta \Phi = -a.\sin \alpha$, and for kinetic energies from 10 eV to 100 eV, the values of $(a)$ are given in table (\ref{table_6}) : 

\begin{table}[h]
\centering
\caption{}
\label{table_6}
\begin{tabular}{|l|l|l|l|l|}
\hline
  $K_{P_{s}}$ (eV) & $\lambda$ $(\text{\AA})$, &  v $(10^{6}\,m.s^{-1})$  & $a (rd)$ & $L_{1/2}$ (cm)\\
\hline
10    & 2.742   & 1.362 & 5.737 $10^{-4}$ & 13.4\\
30    & 1.583   & 2.297 & 9.420 $10^{-4}$ & 22.6\\
50    & 1.226   & 2.965 & 1.218 $10^{-3}$ & 29.2\\
100   & 0.867  & 4.193 & 1.721 $10^{-3}$ & 41.3\\
\hline
\end{tabular}
\end{table}
\vspace{2mm}
\par

With these very small numbers, the measurement of the gravitational phase shift requires a very large number of $(P_{s})$ atoms : one needs a high statistics to sample the small phase shift.

\vspace{2mm}
\par

Unlike atoms interferometers with light standing wave diffraction, the $(P_{s})$ is and remain in the same electromagnetic internal state $(n=0)$ during its path through the interferometer. What is being measured is the rate of $(P_{s})$ decay in the two arms. The only constraint is that its life time is large enough to let it reach the 2 beams recombination region. In the following we will consider ortho-positronium in the ground state with a life time of 142 ns. There will be no external electro-magnetic field involved and hence no splitting of the hyperfine states.

\vspace{2mm}
\par



Aside for surface defects or contamination, the main effect to consider, as limiting the coherence of the diffracted $(P_{s})$ beam and the gravitational phase shift measurement, are the lattice elastic vibrations. At room temperature, the amplitude of lattice elastic vibrations is $\sim$0.1 $\text{\AA}$, hence $\sim5\%$ of the lattice size. This effect will therefore not prevent the interference of de Broglie waves at wave lengths of $\sim2\,\text{\AA}$. By reducing the temperature of the crystal to few kelvins, the optical phonon background can be reduced by a factor $\sim10$.
\vspace{2mm}
\par

If we assume that the supporting structure is a single crystal like in the COW experiment, then the alignment and the parallelism will not be afftected by the elastic vibrations along this support. What may blur the interference is the displacement of the diffracting lattice of the first layer of atoms in Al or Mg, when the $(P_{s})$ atoms are produced by the emission of Surface Plasmons Polaritons. The SPP field is evanescent and decays exponentially over $\sim20$ nm. The SPP are emitted by surface electrons oscillations and the surface thickness where the electrons are captured by the positrons is $\sim\,0.1\,\text{\AA} $.

\vspace{2mm}
\par

The acoustic phonon speed in Al and Mg is $\sim5000 \,m.s^{-1}$ to be compaired to the $1.362 10^{6} \,m.s^{-1}$ of the $(P_{s})$ beam at 10 eV. The time an atom needs to cross a thickness $d$ is $(d/1.362 \,10^{6})$, and during that time, the propagation distance of the elastic elastic wave is $(d/1.362 \,10^{6}) \times 5000$. Hence the distance over which the atom wave can be affected is :

\begin{equation} \label{eq:14}
\begin{aligned}
&(5000 / 1.362 \,10^{6}) \, \times 0.1\, \text{\AA} \sim 3.6\, 10^{-4} \,\text{\AA} \\
\end{aligned}
\end{equation} 

The contribution of these phonons to the measured phase shift at the interferometer exit is :

\begin{equation} \label{eq:14a}
\begin{aligned}
&\lambda = 2.742 \,\text{\AA} \\
&(3.6\, 10^{-4} /  2.742 ) \, \times 2 \pi = 8.2  \,10^{-4} rd >  \Delta \Phi
\end{aligned}
\end{equation} 

Hence the phonons can blur the gravitational phase shift.

\vspace{2mm}
\par
  
If it were not for the decrease in specular reflexion when the kinetic energy grows, it would seem more advantageous to use atoms with a kinetic energy of 100 eV to increase $ L_{1/2}$. The authors who measured the specular reflexion noted that their value is probably lowered by the surface contamination of their sample, and that a reflexion coefficient nearer to 1 is to be expected. If that turn out to be experimentally proven, then the kinetic energy could be increases. But that would not help because the wave length would decrease to $0.867\text{\AA}$ making the phase shift even more difficult to measure within the lattice vibrations background.

\vspace{2mm}
\par

Even at room temperature and with a $(P_{s})$ beam kinetic energy of 10 eV, it possible to measure the phase shift by increasing the path lengh. If the path lengh is increased, for instance $L = 20 \times L_{1/2}$, the maximal phase shift becomes $a = 114.74 \, 10^{-4} rd$ at a cost of dividing the number of $(P_{s})$ atoms in the detectors by a factor 2048. With this loss factor, the maximal phase shift is 14 times larger than the lattice elastic vibrations effect, and hence becomes visible.

\vspace{2mm}
\par

Let's assume that the goal is to measure a difference between the two outputs of the interferometer of $\sim1000$ atoms. Then to measure a maximal phase shift $\Delta \Phi = a\sim 0.01\,rd$ one needs $\sim 2\,10^{5}$ atoms if no atom is lost and if the detector has no jitter in the atom counting. But the $(P_{s})$ decay in flight gives a loss factor 2000 for $a = 114.74 \, 10^{-4} rd$. Then to measure steps of $1/10$ of the maximal phase shift, one needs $10 \times 2000 \times 2\,10^{5} = 4\,10^{9}$ atoms. In real life there are several sources of noise in the counting. So we shall assume an overall efficiency of $10\%$. The specular reflexion of the beam splitters adds a loss factor $\sim10^{-3},10^{-4}$. So the source shall provide $10^{13}\sim 10^{14}$ atoms per measurement. And $100\sim1000$ measurements are performed at each step between $\alpha=0$ to $\alpha=\pi$.

\vspace{2mm}
\par

A classical linac source, with a tungsten target of $\sim$1 mm, water cooled, orthogonal to the 10 MeV electron beam, and followed by a tungsten moderator has a maximal output of $\sim10^8$ $(e^{+})$ per second. So with a classical linac source, one can perform 1 measurement per day. A thin tungsten target at grazing incidence with a magnetic bulb, and no need for target cooling, shall produce $\sim 10^{11}$ $(e^{+})$ per second \cite{rosowskyperez-s}. For detecting the phase shift with a crude measurement, the classical linac source is enough, but to reach a precision similar to the one of neutrons interferometry, like the COW experiment, the thin tungsten target source is required. The classical source experiment can be improved if it is performed at a lower temperature. All the parameters have been taken with a conservative approach.

\vspace{2mm}
\par

An alternative approach is to increase the phase shift by adding locally, near one arm of the interferometer, a large mass. The experiment was perfomed with 84 kg of lead and with a matter interferometer\cite{Asenbaum}. The use non decaying matter atoms allows for 8 m altitude flight which is not practical with $(P_{s})$ atoms, but one could use a much larger mass to increase the phase shift in the $(P_{s})$ interferometer. 

\vspace{2mm}
\par

The loss of coherence due to the lattice vibrations coud be used as a modulator and frequency filter for the $(P_{s})$ beams : Surface Acoustic Wave (SAW) devices are common in electronics and provides all the required toolbox. A specific use of the SAW could be the bunching of diffracted $(P_{s})$ beam to be packaged in short bunches. 

\vspace{2mm}
\par

Aside from pratical effects like the thermal vibrations, one may object that the decay in flight, within each arm of the COW interferometer, provides information about the path of the atoms, and hence erases the interference pattern. But the coherence is not lost : reccording the decays in flight will not provide information about the non decayed atoms path. A similar objection was raised and answered, by Rasel et all \cite{atom-interferometer}, in the case of the standing light wave interferometer.

\vspace{2mm}
\par

The signal collected is the number of $(P_{s})$ atoms after the beam recombination. As with any Mach-Zehnder interferometer, there are two detection directions of interest : one looking at each incoming beam line of flight. When the interference is constructive in one, it shall be destructive in the other. The use of a TEM grid to support the semi-transparent $(P_{s})$ mirrors creates a shadow, and the transmission and reflection coefficients are no the same when the device is entered from one side or the other. Hence the anticorrelation between the 2 detectors shall be observed, but the absolute number of atoms, expected on each detector, shall take into account this effect.

\subsection{The $(P_{s})$ counters}

\begin{figure}[htb]
  \centering
   \includegraphics[trim={2.0cm 8.0cm 0.0cm 2.0cm},clip,scale=0.4]{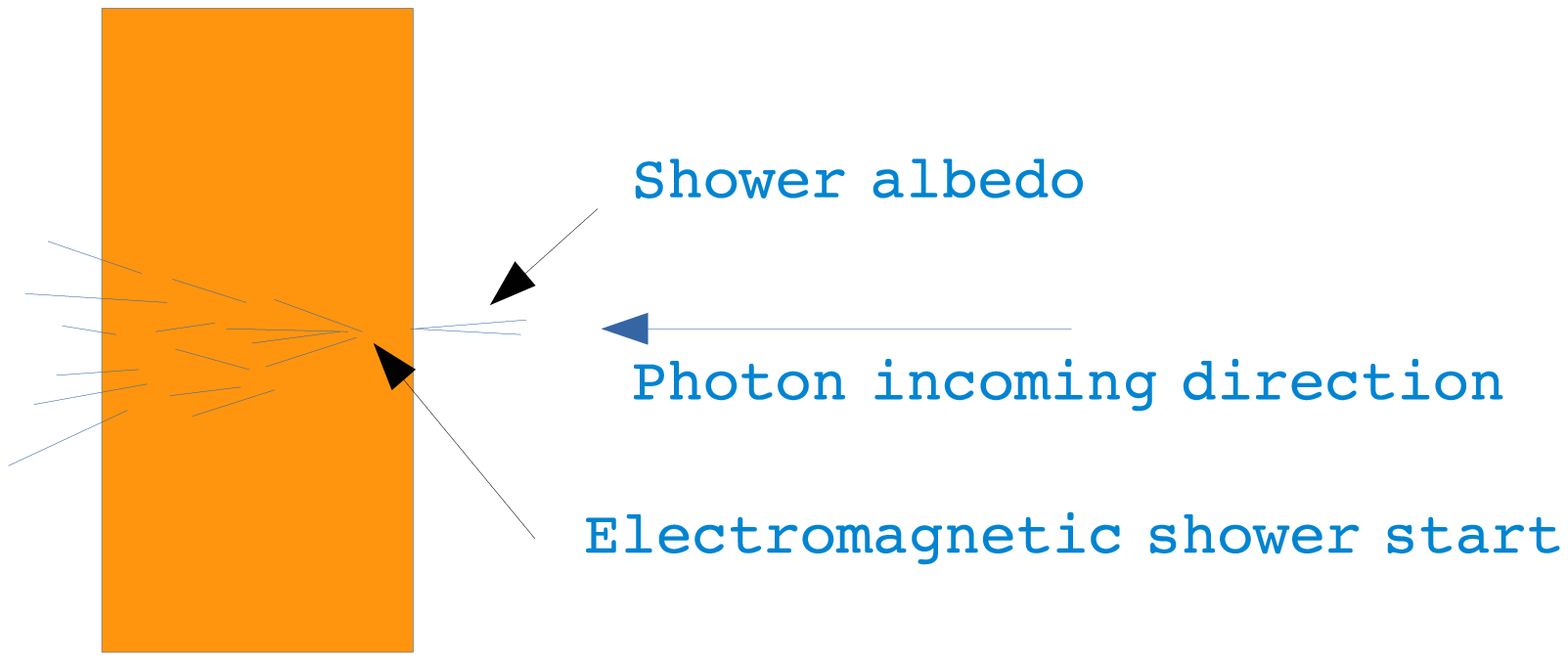} 
  \caption{Electromagnetic shower albedo counter}
  \label{my_fig6}
\end{figure}

\begin{figure}[htb]
  \centering
  \includegraphics[trim={4.0cm 8.0cm 0.0cm 2.0cm},clip,scale=0.4]{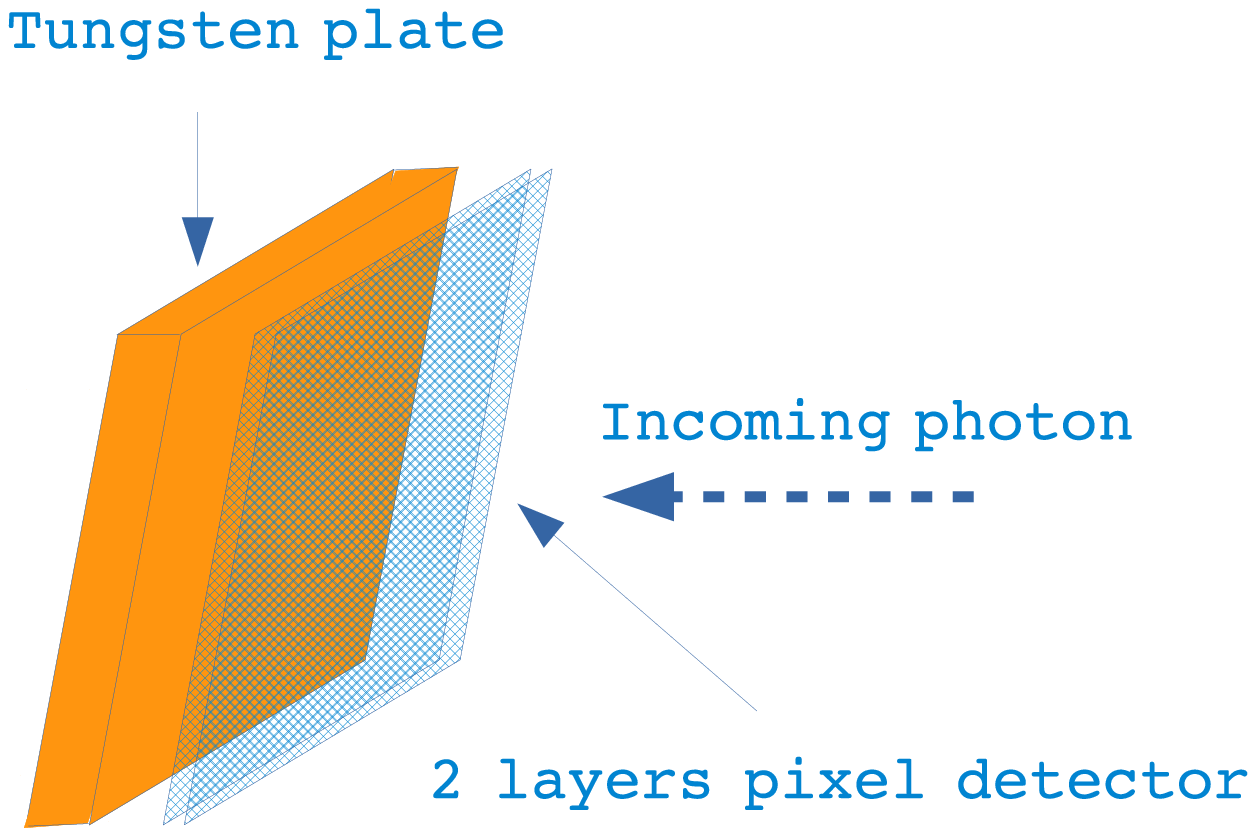} 
  \caption{Electromagnetic shower albedo counter}
  \label{my_fig7}
\end{figure}

A detector will be floaded by photons due to the continuous decay in flight of $(P_{s})$ atoms in the beams. Furthermore to measure an excess of 1000 decays out of $\sim 2\,10^{5}$ requires an accuracy better than $0.5\%$. With each decay producing 1 MeV, it means to distinguish 1 GeV out of 200 GeV. A classical method based on calorimetry is unpractical. But one can use an unconventional path by counting the decays instead of measuring the energy deposited. Lets force the decay channel into back to back 511 keV photons, by flipping the $(P_{s})$ to its singlet state with a 2 Tesla field. Then instead of recording the photons with a calorimeter, one can use the albedo of the electromagnetic shower, figure (\ref{my_fig6}). The albedo is the least fluctuating energy deposited in an electromagnetic shower : it is constant at $\sim5\%$ for any photon energy. So each photon will emit $\sim25$ keV in the albedo. Assuming the internal 2 Tesla magnet is 10 cm diameter and 20 cm long, one can place a 12 faces albedo detector. Each detector is made of a 1 cm thick tungsten plate and a silicon pixel detector facing the tungsten, as illustrated in figure (\ref{my_fig7}). The photon crosses first the silicon detector and then the tungsten. Either it starts it shower in the silicon, or else the shower albedo will hit the silicon. The albedo is emitted backward from the photon line of flight, hence the position can be very precise, $\sim50\mu m$. By counting the hits and dividing by 2, one gets the number of $(P_{s})$ atoms. Since the number of atoms is $\sim 2\,10^{5}$, and each gives 2 photons, with a 20 megapixel silicon detector the occupancy will be $2 \times 2.\,10^{5} / 20. \,10^{6} = 5\%.$

\vspace{2mm}
\par
With a pixel detector efficiency of $99\%$, and a detector made of 2 pixel layers, the detection loss is $10^{-4}$, so the counting precision is much better that the $0.5\%$ required. The use of 2 pixel layers allows the correction of geometrical effects when a photon hit the layer at the junction between 2 pixels. The 2 layers are positioned with an offset of 1/2 a pixel to correct that effect. The magnetic beam orientation does not play any role, but to avoid detection losses by the decay at the exit of the magnetic field, the pixel detectors shall extend farther than the leaking field. Finally, with a cylindrical detector aligned with the $(P_{s})$ line of flight, the decays along this line will not be detected. Hence an endcap double layer pixel detector shall be installed to close the acceptance. Since the 2 photons are emitted back to back, in the endcap, each photon detected contribute to 1 unit in the counting, while in the barel it contributes to 1/2.

\section{Gravity, atoms interferometry, antimatter : controversies}

This experiment will hopefully be done, but there are past and present controversies that will be shortly discussed here.

\vspace{2mm}
\par

In the Quantum Mechanical frame, the phase shift accumulated within two paths, starting at the same altitude, but travelling at different altitudes, is due to the loss of kinetic energy for the path going upward, loss which translates into a difference in momentum, hence in wave length. The loss is equal to the difference in gravitational potential energy. 
By measuring the phase shift and compairing it with the computed one due to gravity, one put to test the weak equivalence principle : the Universal Free Fall.

\vspace{2mm}
\par

In the frame of General Relativity, the path farther from earth feels a weaker gravitational acceleration which produces a gravitational redshift of its wave length and a phase shift with the other path. In that persepective, one put to test the gravitational redshift. This interpretation of the atomic interferometer, proposed in 2011, was rejected but shed light on the concept of time measurement.

\vspace{2mm}
\par

These opposing views were discussed at the \enquote{Rencontres de Moriond 2011} session Weak Equivalent Principle, thursday 24 mars 2011, morning session \cite{wolf1}\cite{Hohensee1}\cite{wolf2}\cite{muller}\cite{Hohensee2}\cite{Hohensee3}\cite{wolf3} for atomic COW experiments using Raman splitters and mirors on standing light waves. After the presentation of the opposite views, a debate took place on the meaning of \textit{time} within each point of view. The audience agreed that the bottom line was to define \enquote{what is a clock ?} \cite{controversy11}. At Moriond no solution was found, and after a long debate .. it was decided to close the session and go skying.

\vspace{2mm}
\par

This question was later solved by defining a clock, that is the condition required to have an object that can tick\cite{controversy12}. The state wave itself is not an observable, and the internal phase of an eigenstate is not a clock. In order to have a clock one has to build a superposition of eigenstates to produce transitions. A stationary state is not a clock, but an unstable one is : what is being measured is a transition rate. In that perspective, the $(P_{s})$ atom in the ground state for electromagnetism cannot be used as a clock, but it can be used for a free fall measurement of the gravitational acceleration $(g)$. 

\vspace{2mm}
\par
While this is true for the classical electromagnetic interaction, it is not so for the Standard model $(e^{+},e^{-})$ annihilation.

\vspace{2mm}
\par
The $(P_{s})$ atom in the electromagnetic ground state (n=0) decays via $(e^{+},e^{-})$ annihilation : it is not an eigenstate. 

\vspace{2mm}
\par

What is being measured is the number of surviving $(P_{s})$ atoms at the exits of the interferometer, hence decay rates. The interferometer then meets the requirement of a clock definition. But since it involves anti-matter, it is not proposed here that it can be used as a redshift measurement. While introducing supergravity, the available extensions of the Standard Model are not experimentally proven. Hence the transitions which are potentially involved in the matter anti-matter annihilation and which relates to gravity are still a matter of theoretical speculations. Compaired to the Standard Model predicted annihilation cross section, these effects, if any, will be very small, and the redshift is also very small. The proposed experiment scope is then limited to the free fall involving anti-matter via interferometry.
\vspace{2mm}
\par

The reference experiment to which the $(P_{s})$ interferometer shall be compared to, is the free fall of an anti-hydrogen atom. Wether such a free fall experiment finds $(g = \bar{g})$ or  $(g \neq \bar{g})$, the phase shift measured in the $(P_{s})$ interferometer with the averaged value of $(g)$ and $(\bar{g})$ shall agree with measured values of these parameters.
\vspace{2mm}
\par
The GBAR experiment, where wave interference does not play any role, shall agree with the gravitational interferometer. If there is a difference between the gravitational interaction of a (anti)matter-(anti)matter system versus a matter-anti-matter system, then the experiment shall see a departure from the computed phase shift, and hence from the number of atoms detected in each detector.

\vspace{2mm}
\par

An other controversial clock argument at Moriond 2011 was the role played by the laser-atom interaction and the effect of the internal states changes during the beam splitting, and at the reflection on standing light waves. This point was adressed in 2014, and it was shown that the laser-atom interaction was not affected by the gravitational redshift\cite{atom-interferometer2}\cite{atom-interferometer1}\cite{atom-interferometer0}. By using $(P_{s})$ atoms in the ground state in a region void of electromagnetic fields (or at least where the fields are neglectable, unlike with lasers), the proposed experiment is not subject of laser atoms effects.

\vspace{2mm}
\par

There are others sources of controversy. The fact that one could contemplate the possibility of anti-gravitation triggered objections \cite{controversy1}. While the above author presents arguments against anti-gravitation, others support it. Answers to the objections to antigravity based on the $K^{0}\bar{K^{0}}$ oscillations, as well as cosmological arguments, and the introduction of the Dirac-Milne universe have been proposed as a theoretical frame for anti-gravitation \cite{controversy3}\cite{controversy5}\cite{controversy6}\cite{controversy7}\cite{controversy8}\cite{controversy9}\cite{controversy10}.

\vspace{2mm}
\par

Antigravity may not exist, and the experiments involved in this search may be futile. But before rejecting them, may be one would contemplate the possibiliy to find an explanation for the galaxy rotation curve with attractive gravity alone, in the frame of General Relativity. As long as the galaxy rotation curve contradicts General Relativity, it is legitimate to question its validity at the edge of existing knowledge with anti-matter.

\section{Acknowledgments}
I wish to thanks Professor Yasunori Yamazaki, from the CUSP collaboration at CERN, and from RIKEN in Japan, with whom I had the pleasure to discuss the paths toward gravity experiments and the CUSP parameters. It is also with the help of Professor Yasunori Yamazaki that I was able to make the first steps in this field 10 years ago, when he supported the proposal to follow the beam on target path to produce $(\bar{H}^{+})$, that Patrice P\'{e}rez and I proposed, and which has become today the GBAR experiment.

\bibliography{cusp-antigravity}{}

\end{document}